\documentclass{article}
\usepackage{epsf}
\newcommand{\be}{\begin{equation}}
\newcommand{\ee}{\end{equation}}
\newcommand{\beq}{\begin{equation}}
\newcommand{\eeq}{\end{equation}}
\newcommand{\ba}{\begin{eqnarray}}
\newcommand{\ea}{\end{eqnarray}}

\newcommand{\id}{i\!\!\not\!\partial}

\newcommand{\ds}{\not\!\! D}

\newcommand{\bea}{\begin{eqnarray}}
\newcommand{\eea}{\end{eqnarray}}

\begin{document}
%%%%%%%%%%%%%%%%%%%%%%%%%%%%%%%%%%%%%%%%%%%%%%%%%%%%%%%%%%%%%%%%%%%
\title{Anomalies in Noncommutative Dipole Field Theories}
\author{D.H.~Correa$^a$\thanks{CONICET} \,,
G.S.~Lozano$^b$\thanks{Associated with CONICET} \, , \\
E.F.~Moreno$^{a\dagger}$ and
F.A.Schaposnik$^a$\thanks{Associated with CICPBA}\\
{\normalsize\it $^a$Departamento de F\'\i sica, Universidad Nacional
de La Plata}\\
{\normalsize\it C.C. 67, 1900 La Plata, Argentina}
\\
{\normalsize\it $^b$Departamento de F\'\i sica, FCEyN, Universidad de
Buenos Aires}\\
{\normalsize\it Pab.1, Ciudad Universitaria,
CP 1428, Buenos Aires,Argentina}
}
%\date{\hfill}
\maketitle

%===================================================================
\begin{abstract}
We study chiral symmetries of fermionic non commutative dipole theories. By using Fujikawa's approach we obtain explicit expressions of the anomalies for
Dirac and chiral fermions in 2 and 4 dimensions.
\end{abstract}
%\tableofcontents

\section{Introduction}

Noncommutative field theories have proven to be a rich nonlocal generalization
of ordinary field theories, with appealing connections with string theories (see
ref.\cite{Ha} for a complete list of references). Apart from noncommutative theories in the Moyal plane, basically defined by
the coordinate commutator algebra $[x_\mu,x_\nu] = i \theta_{\mu\nu}$ (with
$\theta_{\mu\nu}$ the constant antisymmetric matrix measuring
noncommutativity), other nonlocal field theories, known as ``dipole field
theories'' have recently been proposed  \cite{BG}-\cite{DSJ}. In fact,
a key feature of noncommutative field
theories is that they describe the dynamics of dipole-like objects, with
electric  dipole
moment related to their momentum according to \cite{SJ1}-\cite{BS}
\begin{equation}
L^\mu \sim \theta^{\mu \nu}p_\mu
\end{equation}
Dipole theories share this property  but they are
much simpler, their intrinsic dipole vectors $L^\mu$ being constants.

Dipole theories were obtained in \cite{BG} by studying T-duality of twisted
noncommutative gauge theories. They can also be connected with Dp-branes
pinned by a potential related to the B field \cite{DGR}. They are also
interesting theories  {\it per se}: they show signs of CP and CPT violations,
they allow to define sensible noncommutative $SU(N)$ gauge theories and
supersymmetric extensions and they could be used to build up a
noncommutative version of the Standard model \cite{DSJ}.

In  this work we discuss the issue of anomalies in dipole gauge theories. First,
we study the $U(1)$ chiral anomaly for a $U(N)$ dipole gauge theory coupled to
Dirac fermions both in the fundamental (or anti-fundamental) and adjoint
 representations. Then, we consider the case in which fermions have a chiral coupling
 to the gauge field, thus studying gauge anomalies in dipole theories.
Our analysis can be summarized as follows. We first consider a dipole
theory consisting of Dirac fermions coupled to $U(N)$  gauge fields. We
study the conservation of the $U(1)$ chiral current both in $d=2$ and $d=4$
dimensions. Using
a path-integral approach, we show that there is a chiral anomaly which,
for fermions in the ``fundamental'' or ``antifundamental'' representations,
could be trivially inferred from the corresponding result in ordinary
(commutative) theories by an appropriate renaming of the gauge fields. In
contrast, when fermions are considered in the ``adjoint'' a genuine new and
non trivial result is obtained. A distinctive feature should be signaled in
comparison with Moyal noncommutative theories: for vanishing dipole length one
recovers the trivial result to be expected for fermions in the adjoint in
ordinary theories, in contrast with what  it happens for Moyal theories due to
the IR/UV mixing \cite{Ard}-\cite{Martin}.
We also consider the issue of gauge anomalies in dipole theories with Weyl
fermions. Using the Fujikawa approach, we analyze the covariant form of the
gauge current anomaly finding in this case that the $U(1)$ and $SU(N)$
contributions have a mixed dependence on the dipole length so that the result
cannot be related, even for the fundamental or antifundamental
representations,  to that in commutative theories by a renaming of the  gauge
fields.

\section{The chiral anomaly}

Given two functions $\phi(x)$, $\chi(x)$, defined in  $d$ dimensional Euclidean
space-time,  the ``dipole star
product''
is defined as
\be
\phi(x) * \chi(x) \equiv \phi( x - \frac{L_\chi}{2}) \chi (x +
\frac{L_\phi}{2})
\label{1}
\ee
where $L_\phi$ ($L_\phi = (L_\phi^\mu)$,  $\mu = 1,2,\ldots ,d$) is a
constant
vector measuring the ``dipole length'' associated with $\phi$. This
noncommutative product is  associative provided that
\be
L_{\phi*\chi} = L_\phi + L_\chi \;\;.
\label{2}
\ee
Given $N$ functions $\phi^a(x)$ ($a=1,2,\ldots,N$) satisfying
\be
\sum_{a=1}^{N} L_\phi^a   = 0
 \;\;,
\label{3}
\ee
one can use integration over $R^d$ to define a trace-like operation
\be
\int d^dx \, \phi^1 * \phi^2 * \ldots * \phi^{N-1}* \phi^N =
\int d^dx \, \phi^N * \phi^1 * \phi^2 * \ldots * \phi^{N-1} \;\;.
\label{4}
\ee
Calling $\phi^\dagger$ the complex (hermitian) conjugate  of $\phi$ one
can see that
\be
\left( \phi^\dagger * \phi \right) = \left( \phi^\dagger * \phi
\right)^\dagger
\; \Leftrightarrow \;
 L_\phi = - L_{\phi^\dagger}
 \;\;.
\label{5}
\ee
For  a fermionic  field $\psi(x)$, one then has
\be
L_{\bar \psi} = - L_\psi
 \;\;,
\label{8}
\ee
while real fields should have zero dipole length.
In particular,
for  gauge fields $A_\mu$, one necessarily has
\be
L_{A_\mu} = 0
 \;\;.
\label{6}
\ee
This in turn  implies that the field strength has the form it takes
for ordinary products since, being the dipole length zero,
$A_\mu(x)*A_\nu(x) = A_\mu(x) A_\nu(x)$.

We  shall  consider in this section the case of  Dirac fermions $\psi$
coupled to $U(N)$ gauge fields, $A_\mu(x) = A_\mu^a(x) t^a$  with $t^a$
the $U(N)$ generators ($a=0,1,\ldots,N^2-1$),   normalized
according to
\be
{\rm tr^g} (t^at^b) = \frac{1}{2} \delta^{ab} \;\;.
\label{gen}
\ee

 Under local gauge
transformations, $A_\mu(x)$ changes as
\ba
 A_\mu(x) &\to& A^g_\mu(x) = g(x) * A_\mu(x) * g^{-1}(x)
- \frac{i}{e} \partial_\mu g(x) * g^{-1}(x) \nonumber\\
&& \;\;\;\;\; \;\;\;\;\; =
g(x)  A_\mu(x)  g^{-1}(x)
- \frac{i}{e} \partial_\mu g(x)  g^{-1}(x)
\label{A}
\ea
with $g(x) \in U(N)$. Concerning Dirac fermions, one can define,
as for the ordinary Moyal product,  gauge
transformations in a
``fundamental'' representation,
\ba
 \psi(x) &\to& \psi^g(x) = g(x) * \psi(x)
  \nonumber\\
  ~
  \nonumber\\
  \bar \psi(x) &\to& \bar \psi^g(x) = \bar \psi(x) * g^{-1}(x)
 \;\;.
  \label{12}
\ea
Alternatively, one can consider an ``antifundamental'' representation
with fer\-mions transforming as
\ba
 \psi(x) &\to& \psi^g(x) = \psi(x) * g^{-1}(x)
  \nonumber\\
  ~
  \nonumber\\
  \bar \psi(x) &\to& \bar \psi^g(x) = g(x) * \bar \psi(x)
 \;\;.
  \label{12af}
\ea
Finally, there is the possibility of fermions in the ``adjoint''
representation.
 Under gauge transformations, they change according to
\be
\psi(x) \to \psi^g(x) = g(x) * \psi(x) * g^{-1}(x)
 \label{13}
\ee
with $\psi = (\psi_i^j)$, $i,j = 1,2, \ldots, N$.
The corresponding covariant derivatives  take the form
\begin{eqnarray}
D^f_\mu[A]\psi(x) \!\!&=&
\!\!
 i\partial_\mu \psi(x)  + e A_\mu(x)*\psi(x) =
 i \partial_\mu \psi(x)  + e A_\mu(x - \frac{L_\psi}{2})\psi(x)
\nonumber\\
  D^{\bar f}_\mu[A]\psi(x) \!\! &=&
 \!\!
i\partial_\mu \psi(x) - e \psi(x) *A_\mu(x) =
i\partial_\mu \psi(x) - e \psi(x) A_\mu(x + \frac{L_\psi}{2})
\nonumber\\
D^{ad}_\mu[A]\psi(x)\!\!  &=&  \!\! i\partial_\mu \psi(x)  + e[
A_\mu(x),\psi(x)]
\nonumber\\
&=& i \partial_\mu \psi(x)  + e A_\mu(x - \frac{L_\psi}{2})\psi(x) -
e\psi(x) A_\mu(x + \frac{L_\psi}{2})
 \;\;.
\label{9}
\end{eqnarray}
For each one of the previous cases, one can construct a gauge invariant  Dirac
action,
\be
S[\bar \psi,\psi,A] = \int d^dx  \bar \psi(x) *  \gamma^\mu D_\mu[A]
\psi(x)
\label{10}
\ee

\be
S[\bar \psi^g,\psi^g,A^g] = S[\bar \psi,\psi,A]
 \;\;,
\label{14}
\ee
where a trace
over the gauge group indices is implicit in the definition of the
action.

At the classical level, the action
(\ref{10}) is also invariant under global
$U(1)$ {\em chiral} rotations. For all three representations
the global chiral rotation (with infinitesimal
 phase $\delta \phi$) takes the same form,
\ba
& & \psi(x) \to \psi^{\delta\phi} = \exp\left(\gamma_5\delta\phi\right)
\psi(x)
\nonumber\\
& & \bar \psi(x) \to \bar \psi^{\delta\phi} =
 \bar \psi(x)   \exp\left(\gamma_5
\delta\phi\right)  \nonumber\\
  & & S[\bar \psi^{\delta\phi},\psi^{\delta\phi},A] = S[\bar
\psi,\psi,A]
 \;\;.
  \label{15}
\ea

At the quantum level, the fermionic  effective action $S_{eff}[A]$ is
defined by
\be
\exp\left(-S_{eff}[A]\right) = \int D\bar\psi D \psi
\exp\left(-S[\bar \psi,\psi,A] \right) \equiv \det \left(\ds [A]
\right)
\label{11} \;\;.
\ee
As in the ordinary case, a path-integral measure for Dirac
fermions can be defined so that the resulting
effective action is gauge
invariant,
\be
S_{eff}[A^g] = S_{eff}[A]
 \;\;.
\label{16}
\ee

Nevertheless,
the  chiral invariance  of the  quantum effective action  is not {\em a priori}
guaranteed, that is, an {\em anomaly} could arise.

In order to analyze this issue, we shall follow Fujikawa's approach (see
\cite{Fujikawa}-\cite{GMSS} for details). In particular, we shall focus on the {\em adjoint representation} since as we shall discuss later, the results for the other two cases (fundamental and antifundamental representations) can be trivially derived from the corresponding results in ordinary commutative field theories.

Expanding the fermion field in
an appropriate basis $\{\varphi_n(x)\}$ with Grassman coefficients $c_n$
\be
\psi(x) = \sum c_n \varphi_n(x) \, , \;\;\;\;  \bar \psi(x) = \sum \bar
c_n
\varphi_n(x)^\dagger
 \;\;,
\label{18}
\ee
 the path-integral measure is formally
defined
as
\be
D\bar \psi D \psi \equiv \Pi d\bar c_n dc_n
\label{17} \;\;.
\ee
One then promotes
the global chiral rotation
(\ref{15}) to a local change of variables in the path-integral.
%%%%%%%%%%%%%%%%%%%%%%%%%%%%%%%%%%%%%%%%%%%%%%%%%%%%%%%%%%%%%
In noncommutative space, there are various possibilities to do this. One
can consider the following three

\ba
& & \psi(x) \to \psi^{\delta\phi} = \exp\left(\gamma_5\delta\phi(x)\right)
*\psi(x)
\nonumber\\
& & \bar \psi(x) \to \bar \psi^{\delta\phi} =
 \bar \psi(x) *  \exp\left(\gamma_5
\delta\phi(x)\right)   
  \label{15ca}
\ea
 
\ba
& & \psi(x) \to \psi^{\delta\phi} = 
\psi(x) * \exp\left(\gamma_5\delta\phi(x)\right)
\nonumber\\
& & \bar \psi(x) \to \bar \psi^{\delta\phi} =
  \exp\left(\gamma_5
\delta\phi(x)\right)   *  \bar \psi(x) 
  \label{15ca2}
  \ea
\ba
& & \psi(x) \to \psi^{\delta \phi} (x) =
 \exp\left(\gamma_5[\delta\phi(x),\;\;]_*\right) \psi(x)
\nonumber\\
& & \bar \psi(x) \to \bar \psi^{\delta \phi} (x) =
 \bar \psi(x)   \exp\left([\;\;,\delta\phi(x)]_*\gamma_5
\right)
 \;\;.
\label{15ad}
\ea
To each of these transformations it corresponds the following classically 
conserved
currents

\[
j_\mu^{5\,(1)}(x) = i \left(\gamma_\mu\gamma_5 \right)^{\alpha \beta} {\rm tr^g}
\left( \psi^\beta (x) * \bar \psi^\alpha (x)
\right)
\]

\[
j_\mu^{5\,(2)}(x) = i \left(\gamma_\mu\gamma_5 \right)^{\alpha \beta} {\rm tr^g}
\left(   \bar \psi^\alpha (x) * \psi^\beta (x)
\right)
\]

 \be
j_\mu^{5\,(3)}(x) = i   \left(\gamma_\mu \gamma_5\right)^{\alpha\beta} {\rm
tr^g}
\!\!\left(
\psi^\beta(x)*\bar \psi^{\alpha}(x) + \bar \psi^{\alpha}(x)
*\psi^\beta(x)
\right)
\label{20ad}
\ee
respectively. Notice that, in view of the definition of the dipole product, 
\[j_\mu^{5\,(1)}(x) = -j_\mu^{5\,(2)}(x+L)
\]
 and then in the 
commutative ($L \to 0$) limit both
currents coincide. Concerning $j_\mu^{5\,(3)}(x)$, it is the current which
naturally couples in the adjoint to an axial $U(1)$ vector field and it vanishes
in the $L \to 0$ limit.
 
%%%%%%%%%%%%%%%%%%%%%%
These  change of  variables could in principle give rise to a non-trivial
Jacobian $J$ in the path-integral measure,
\be
  D\bar \psi^{\delta \phi} D \psi^{\delta \phi} = J^2
D\bar \psi D \psi 
 \;\;.
\label{22}
\ee
 Using
that
$\delta Z/\delta \phi(x) = 0$
,
one obtains the anomaly equation
\be
\langle \partial_\mu j_\mu^5 \rangle = {\cal A}(x)
\label{19}
\ee
Let us concentrate for definiteness in the case
 of current $j_\mu^{5\,(3)}(x)$. The
 anomaly ${\cal A}$ appearing in
 the r.h.s. of (\ref{19}),   is related to the Fujikawa Jacobian
through
 the formula
\be
{\cal A}(x) = \frac{\delta \log J^2}{\delta \phi(x)}
 \;\;.
\label{21}
\ee
It takes the form
\be
{\cal A} = 4 {\rm tr^g}\sum_n \gamma_5^{\alpha \beta}
\left(\varphi_n^\alpha *
 {\varphi_n^\dagger}^\beta  - {\varphi_n^\dagger}^\alpha *
 \varphi_n^\beta
\right)
 \;\;.
\label{23ad}
\ee
Expression (\ref{23ad}) is ill defined  and has to be appropriately regulated. A
gauge-invariant result can be obtained by using as a basis 
the set of  eigenfunctions of the Dirac operator $\ds$ (in the appropriate
representation) and then introducing a $\ds$-based
heat-kernel regularization so
that one ends with a finite answer ${\cal A}_{reg}$
\ba
{\cal A}_{reg}^{(3)}(x) &=& 4 \lim_{t \to 0} {\rm tr^g} \sum_n \gamma_5^{\alpha
\beta}
\left(\left(\exp\left(-t \gamma_\mu\gamma_\nu
D_\mu^{ad}D_\nu^{ad}\right)
\varphi_n\right)^\alpha *
 {\varphi_n^\dagger}^\beta  \right.\nonumber\\
 && \left.  - {\varphi_n^\dagger}^\alpha *
 (\left(\exp\left(-t \gamma_\mu\gamma_\nu D_\mu^{ad}D_\nu^{ad}\right)
\varphi_n\right)^\beta
\right) \;\;.
\label{23adi}
\ea

The final expression of the anomaly depends on the space-time dimensionality {\em d}.
Indeed, for a 2-dimensional (4-dimensional) theory only the lineal (quadratic)
term in an  expansion on $t$ contributes to the
result. A standard (but tedious) calculation leads to,
\ba
&&{\cal A}^{(3)\,(d=2)}_{reg}(x) =\frac{eN}{2\pi}\,\sqrt{\frac{N}{2}}\,\varepsilon_{\mu\nu}
 \left(F^0_{\mu\nu}(x - L) + F_{\mu\nu}^0 (x + L) - 2F_{\mu\nu}^0 (x)\right)
\label{ano2ad}
\\
&&{\cal A}^{(3)\,(d=4)}_{reg}(x) =\frac{e^2N}{8\pi^2}\left[
F_{\mu\nu}^a (x+L) \tilde F_{\mu\nu}^a (x+L) -
F_{\mu\nu}^a (x-L) \tilde F_{\mu\nu}^a (x-L)\right.
\nonumber\\ && ~~~~~~~~\left. -
2\left(F_{\mu\nu}^0 (x+L) -
 F_{\mu\nu}^0 (x-L)\right)\tilde F_{\mu\nu}^0 (x)\right] \;\;.
 \label{ecuacioncita4}
 \ea
with $\tilde F_{\mu\nu}=\frac{1}{2} \varepsilon_{\mu \nu \alpha \beta}
 F_{\alpha \beta}$.

An analogous calculation can be performed for the anomaly associated
to currents $j_\mu^{5\,(1)}$ and $j_\mu^{5\,(2)}$.  One gets
\ba
&&{\cal A}^{(1)\,(d=2)}_{reg}(x) =\frac{eN}{2\pi}\,\sqrt{\frac{N}{2}}\,
\varepsilon_{\mu\nu}
 \left( F_{\mu\nu}^0 (x + L) - F_{\mu\nu}^0 (x)\right)
\label{ano2ad1}
\\
&&{\cal A}^{(1)\,(d=4)}_{reg}(x) =\frac{e^2N}{8\pi^2}\left[
F_{\mu\nu}^a (x+L) \tilde F_{\mu\nu}^a (x+L) +
F_{\mu\nu}^a (x) \tilde F_{\mu\nu}^a (x)\right.
\nonumber\\ && ~~~~~~~~\left. -
2 F_{\mu\nu}^0 (x+L) 
 \tilde F_{\mu\nu}^0 (x)\right] \;\;.
 \label{1ecuacioncita4}
 \ea
One gets similar results for the case of current  $j_\mu^{5\,(2)}$.

It is instructive to compare these result with those
arising in the commutative case and with those
corresponding to noncommutative theories defined using
the Moyal product. Taking the limit $L \to 0$ in
(31)-(32) we see that the anomaly vanishes in
agreement with the trivial result one should expect in
ordinary space since $j_\mu^{5\,(3)} =0$ in that case. 
This should be contrasted with the
non-trivial answer one finds for the anomaly in the
$\theta \to 0$ (i.e. commutative) limit of Moyal
noncommutative theories,  as a result of the
well-understood  IR/UV mixing taking place in this
case. Concerning   currents $j_\mu^{5\,(1)}$
and $j_\mu^{5\,(2)}$ the corresponding anomalies vanish in the $L \to 0$ 
limit for  $d=2$
but give a nontrivial result for $d=4$, which coincide with the standard
commutative results.

Let us end this section by briefly describing  the dipole-product
anomaly calculation for the case of the fundamental
and anti-fundamental representations. In principle one could follow the
same steps than for the adjoint representation. Nevertheless we can obtain the results
directly by noticing that by renaming the field
$ A_\mu(x -L/2) = a_\mu(x)$ and $A_\mu(x + L/2) = a_\mu(x)$ for the fundamental and
antifundamental representations the theories are formally identical to the
ordinary commutative field theories. For the fundamental representation
one obtains
\be
\langle \partial_\mu j_\mu^5 \rangle = {\cal A}^f(x)
\ee
where
\be
j_\mu^5 = i \left( \gamma_\mu\gamma_5\right)^{\alpha \beta}
\psi(x)^\beta * \bar \psi(x)^\alpha
\ee
and
\be
{\cal A}^{f(d=2)}_{reg}(x)=
-\frac{e\sqrt N}{2\sqrt 2 \pi}\varepsilon_{\mu\nu}
 F_{\mu\nu}^0 (x )
 \label{anoyprimasi2}
\ee
\be
{\cal A}^{f(d=4)}_{reg}(x)=\frac{e^2}{8\pi^2}   \tilde F^a_{\mu\nu} (x) F^a_{\mu\nu} (x) \;\;.
 \label{anoyprimasi4}
\ee
%%%%%%%%%%%%%%%%%%%%%%%
%%%%%%%%%%%%%%%%%%%%%%%
\section{Gauge anomalies in   theories with Weyl fer\-mions}
In this section we shall analyze the case of Weyl fermions. As before,
the results in the fundamental and antifundamental representations are trivially
obtained by  redefinitions of the gauge fields. We shall therefore consider 
the case of fermions in the adjoint representation. The action for this theory is
\be
S[\bar \psi,\psi,A] = \int d^dx\; \bar \psi(x)* D_+ \psi(x) + S_{g}[A] \;\;.
\label{10ch}
\ee
In this case, the Dirac operator $D_+$ reads
\be
D_{+}= \id + e \gamma^{\mu} P_{+} \;[A_{\mu},\cdot\;]_{*}
\label{c-2}
\ee
where
\be
P_\pm = \frac{1}{2} (1 \pm \gamma_{5}) \;\;.
\ee
and  $S_{g}[A]$ denotes  the gauge fields action.

Notice that the chiral Dirac operator $D_+$ is not Hermitian. In
fact, its adjoint is given by
\be D_+^{\dagger} = D_- = \id + e \gamma^{\mu} P_{-}
\;[A_{\mu},\cdot\;]_{*}  \;\;. \label{c-3.1}
\ee
The action is invariant under the following chiral gauge
transformations
\begin{eqnarray}
\psi &\to& \psi' \; =\;  e^{i P_{+} [\alpha, \cdot]}\; \psi \; =
\;
P_{+} e^{i\alpha} * \psi * e^{-i \alpha} + P_{-}\psi \nonumber\\
{\bar \psi} &\to& {\bar \psi}' \; =\; {\bar \psi}\; e^{-i P_{-}
[\cdot, \alpha]} \; = \; e^{i\alpha} * {\bar \psi} * e^{-i \alpha}
P_{-}+ \bar \psi P_{+}
\nonumber \\
A_{\mu} &\to&  A_{\mu}' \; =\; e^{i\alpha} * A_{\mu} *  e^{-i
\alpha} + \frac{i}{e}\; e^{i\alpha}*  \partial_{\mu} e^{-i \alpha} \;\;.
\label{c-4}
\end{eqnarray}
The current associated to this symmetry is classically conserved
but due to the chiral nature of the transformation the
conservation is anomalous. The current has two contributions, one
from the matter fields and other from the gauge fields. Only the
matter contribution to the current
\be
\left(j_{m}^\mu\right)_{ab} = \left({\bar \psi}^\alpha_{ac}* \psi^\beta_{cb}
+ {\psi}^\beta_{ac}* {\bar \psi}^\alpha_{cb}\right) (\gamma^{\mu} P_+)^{\alpha\beta}
\label{c-5}
\ee
is anomalous so we will concentrate on it. At the classical level it satisfies a covariant conservation law,%
\be
{\cal D}_{\mu} j_{m}^{\mu} = \partial_{\mu} j_{m}^{\mu} - i e
[A_{\mu}, j_{m}^{\mu} ] = 0 \;\;. \label{c-6}
\ee
We shall discuss the covariant form of the gauge anomaly
adopting Fujikawa recipe \cite{Fujikawa}-\cite{Martin} which uses as basis
for fermion fields the  eigenfunctions of the Hermitian
operators $ D_+ D_-$ and $D_- D_+ $. That is, one
considers the eigenvalue problems
\ba
 D_- D_+\varphi_n &=& \lambda_n^2 \varphi_n\nonumber\\
 D_+ D_-\phi_n &=& \lambda_n^2 \phi_n\nonumber\\
 \phi_n = \frac{1}{\lambda_n}D_+ \varphi_n \; \;{\rm if}\;\; \lambda_n \ne 0
 \;&{\rm and}&\; D_+ \varphi_n = 0\;\; {\rm if}\;\; \lambda_n = 0\nonumber\\
 \varphi_n = \frac{1}{\lambda_n}D_- \phi_n \; \;{\rm if}\;\; \lambda_n \ne 0
 \;&{\rm and}&\; D_- \phi_n = 0\;\; {\rm if}\;\; \lambda_n = 0
 \label{dos}
 \ea
and expands the fermion variables in the form
\be
\psi(x) = \sum c_n \varphi_n(x) \, , \;\;\;\;  \bar \psi(x) = \sum \bar c_n
\phi_n^\dagger \;\;.
\label{18ch}
\ee
Again, the path-integral measure is defined as in (\ref{17})
\be
D\bar \psi D \psi \equiv \Pi d\bar c_n dc_n \;\;.
\ee
Performing a chiral gauge transformation of fermion variables, we obtain
for the anomaly
\be
\left\langle \left({\cal D}_{\mu} j_{m}^{\mu}\right)_{ab}\right\rangle
=-\frac{\delta \log({\bar J} J)}{\delta\alpha_{ba}}= {\cal A}_{ab}(x)
\ee
with $J$ and $\bar J$ being the Jacobians associated to the chiral gauge transformation
of $\psi$ and $\bar \psi$. Since the chiral Dirac operators are 
non-Hermitian, those Jacobians
are different and must be computed independently.
After standard computation we have
\be {\cal A}(x) = -2 i \; {\rm tr} \left( \sum_n\; P_+\; \left[
\varphi_n(x), \varphi^{\dagger}_n(x) \right]  - \sum_n\; P_-\;
\left[ \phi_n(x), \phi^{\dagger}_n(x) \right]\right) \label{c-9}
\ee
where the trace is taken only over the Dirac indices.
This quantity is ill defined and has to be regularized. A gauge
covariant regularized expression is given by
\begin{eqnarray}
{\cal A}_{reg}(x) &=& -2 i \;\lim_{t\to 0} {\rm tr} \left(
\sum_n\; P_+\; \left[ e^{- t D_-D_+} \varphi_n(x),
\varphi^{\dagger}_n(x)
\right] - \right. \nonumber \\
&& \left. \sum_n\; P_-\; \left[e^{- t D_+D_-}\phi_n(x),
\phi^{\dagger}_n(x) \right]\right) \;\;.  \label{c-10}
\end{eqnarray}
Changing the eigenstate basis to plane-wave functions and using
the identity
\be P_+ \;e^{- t D_-D_+} - P_-\; e^{- t D_+D_-} = \gamma^{5}\;
e^{- t {\not D} } \label{c-11} \ee
where $\ds$ is the standard Dirac operator
\be \ds = \id + e \gamma^{\mu}\;[ A_{\mu}, \cdot\; ]
\label{c-12}\ee
we obtain 
\begin{equation}
{\cal A}_{reg}(x) = -2i \;\lim_{t\to 0} {\rm tr}
\sum_a\;\int \frac{d^d k}{(2\pi)^d} \left(\gamma^{d+1}\;  \left[
e^{- t {\not D} } e^{ik\cdot x}t^a , e^{-ik\cdot x} t^a \right]
\right) \label{c-13}
\end{equation}
and computing this expression for the cases of $d=2$ and $d=4$, we obtain the following
chiral gauge anomalies
\begin{eqnarray}
d=2 && \hspace{.4cm} {\cal A}_{reg} = i N \frac{e}{4\pi} \;
\varepsilon^{\mu \nu} \; \left(F^0_{\mu \nu}(x-L)t^0 + F^0_{\mu
\nu}(x+L)t^0 - 2 F^a_{\mu \nu}(x)t^a \right) \nonumber\\
d=4 && \hspace{.4cm} {\cal A}_{reg} = i \sqrt{2N} \frac{e^2}{32\pi^2}
\; \left( (F_{\mu \nu}^a(x-L){\tilde F}^{\mu \nu a}(x-L) -
\right. \nonumber\\
&& \hspace{-1.0cm} \left. F_{\mu \nu}^a(x+L){\tilde F}^{\mu \nu
a}(x+L)) t^0 - 2 (F^0_{\mu \nu}(x-L) - F^0_{\mu \nu}(x+L)) {\tilde
F}^{\mu \nu a}(x) t^a \right) \label{c-14}
\end{eqnarray}
Notice that unlike the ordinary commutative case and the noncommutative Moyal case,
there is non-vanishing covariant anomaly in $d=4$ \cite{BoS},\cite{Martin}.
Again, we would like to stress that in the $L \to 0$ limit we recover the result
for the anomalies in the commutative theory because of the non-existence of the
IR/UV mixing in noncommutative dipole field theories.

As we stated in the introduction, dipole theories are
not only interesting in connection with string theory but also because of
their possible relevance as field theories in particle physics. In this
respect, it has been shown that they show signs of CP and CPT violations,
the Lorentz symmetry breaking is soften and the possibility of constructing
a dipole version of the Standard Model more attainable
\cite{DSJ}. In view of this, the issue of chiral and gauge anomalies in dipole
theories is a relevant question, which has been discussed in this work.
Using a path-integral approach, we have computed chiral and gauge anomalies
in $U(N)$ dipole fermion theories. We have found non-trivial results which,
in view of the connection between the anomaly and topological densities pose
in turn the problem of studying topological solutions. We hope to discuss
this aspect in future works.

\noindent\underline{Acknowledgements}: This work  partially
supported  by UNLP, UBA, CICBA, CONICET (PIP 4330/96), ANPCYT (PICT
03-05179), Argentina.  G.S.L. and E.F.M. are partially supported
by Fundaci\'on Antorchas, Argentina. F.A.S wishes to acknowledge 
P.Forgacs and H.Giacomini
and the Laboratoire de Math\'e\-ma\-tiques et Physique
Th\'eorique de l'Universit\'e de Tours for their
hospitality during part of this work.

%\newpage

\end{document}